\documentclass[12pt]{article}
\usepackage{epsfig}

\parskip=\medskipamount 
plus.3em minus.5em \abovedisplayshortskip=.5em plus.2em minus.4em
\renewcommand{\thesection}{\arabic{section}}

\catcode`@=11

\@addtoreset{equation}{section} \@addtoreset{equation}{subsection}
\def\theequation{\ifnum\value{section}=0 \arabic{equation}\ignorespaces
\else \ifnum\value{section}=-1 A.\arabic{equation}\ignorespaces
\else \ifnum\value{subsection}=0
\thesection.\arabic{equation}\ignorespaces \else
\thesection.\arabic{subsection}.\arabic{equation}\ignorespaces
                             \fi
                        \fi
                   \fi}

{\catcode`\'=\active \def'{{}^\bgroup\prim@s}}

\catcode`@=12

\newcommand{\bq}{\begin{equation}}
\newcommand{\be}{\begin{equation}}
\newcommand{\fq}{\end{equation}}
\newcommand{\ee}{\end{equation}}
\newcommand{\bqr}{\begin{eqnarray}}
\newcommand{\beqs}{\begin{eqnarray}}
\newcommand{\fqr}{\end{eqnarray}}
\newcommand{\eeqs}{\end{eqnarray}}

\newcommand{\rf}[1]{(\ref{#1})}

\def\on#1#2{{\buildrel{\mkern2.5mu#1\mkern-2.5mu}\over{#2}}}
\def\dt#1{\on{\hbox{\bf .}}{#1}}                

\def\bop#1{\setbox0=\hbox{$#1M$}\mkern1.5mu
    \vbox{\hrule height0pt depth.04\ht0
    \hbox{\vrule width.04\ht0 height.9\ht0 \kern.9\ht0
    \vrule width.04\ht0}\hrule height.04\ht0}\mkern1.5mu}

\begin{document}
\thispagestyle{empty}

\begin{flushright}
\begin{tabular}{l}
hep-th/0503213 \\
\end{tabular}
\end{flushright}

\vskip .6in
\begin{center}

{\bf  Computational Derivation to Zeta Zeros and Prime Numbers}

\vskip .6in

{\bf Gordon Chalmers}
\\[5mm]

{e-mail: gordon@quartz.shango.com}

\vskip .5in minus .2in

{\bf Abstract}

\end{center}

A route to the derivation of the numbers $s$ to the transcendental
equation $\zeta(s)=0$ is presented.  The solutions to this equation
require the solving of a geodesic flow in an infinite  
dimensional manifold.  These solutions enable one approach to a 
formula generating the prime numbers.

\setcounter{page}{0}
\newpage
\setcounter{footnote}{0}

The derivation of prime numbers has been a problem of longstanding
interest for many years \cite{Zeta}.  This desire has motivated the
solution to the zeros of the Riemann zeta function, $\zeta(s)=\sum n^{-s}$.  
The longstanding conjecture is that these zeros all lay on the imaginary 
axis at $s={1\over 2}+it$.  This conjecture has been publicly examined 
for an approximate $10^{20}$ zeros.  These zeros and the methods used to 
find them are of importance for various reasons, not only for the 
determination of prime numbers.  

In this presentation a means to obtain these zeros is generated, based 
on work in \cite{ChalmersPoly} and \cite{ChalmersGeoDiff}.  The metrics 
on an underlying infinite dimensional geometry associated with the zeta 
function generate an algorithm to finding the locations of $\zeta(s)=0$.  
In one approach, the geodesic flow allows a formula (possibly transcendental) 
to be obtained; the solution of which generates the zeros.  In another 
approach, a differential equation that the zeta function satisfies is 
used to develop another equation that generates the same zeros.  

The Riemann zeta function is defined by

\bqr
\zeta(s)=\sum_{n=1}^\infty {1\over n^{s}} \ ,
\fqr
and admits the product form,

\bqr
\zeta(s)= \zeta(0) \prod_{n=1}^\infty \bigl(1-{s\over \rho_n}\bigr) .
\fqr
By definition $\rho_n$ label all of the zeros of the function; it is
conjectured that all of these zeros lay on the real axis located at
$s={1/2}+i t$.  The zeta function satisfies a number of properties,
including a reflection identity,

\bqr
\zeta(s)=\Gamma(s)\Gamma(-s) (2\pi)^{s-1} 2 \sin(s\pi/2) \zeta(1-s)
\fqr
with values at,

\bqr
\zeta(1-2n)=(-1)^{2n-1} {B_{2n}/2n} \qquad \zeta(-2n)=0 ,
\fqr
and,
\bqr
\zeta(2n) = {(2\pi)^{2n} (-1)^{n+1} B_{2n} \over 2\cdot \Gamma(2n+1)} \ .
\fqr
The values $\rho_j$ specify all of the zeros of the zeta function; they
generate the prime numbers via the original work of Riemann and others 
including von Mangoldt.

Two methods are presented here to the solution of the parameters $\rho_j$.
One method involves solving an infinite number of polynomial equations for
an infinite number of variables.  The second method requires solving a
differential equation with an infinite number of derivatives.

\vskip .2in
\noindent {\it Algebraic Approach}

Consider the value of the zeta function at the points $s=1-2n$.  These
values generate an infinite number of polynomial equations with $n$
ranging from $0$ to $\infty$,

\bqr
(-1)^{2n-1} {B_{2n}/2n} \zeta^{-1} (0) = \prod_j \bigl(1-(1-2n)\gamma_j\bigr)
\fqr
\bqr
= 1 - (1-2n) \sum_j \gamma_j + (1-2n)^2 \sum_{i,j} \gamma_i \gamma_j
 + \ldots \ ,
\label{polynomial}
\fqr
with the coefficients of the symmetric products being $(1-2n)^p$.  The 
gamma parameters are the zeros, $\gamma_i=1/\rho_i$, with $\zeta(\rho_i) 
=0$.  Set the left hand side of the equation to be variables,

\bqr
(-1)^{2n-1} {B_{2n}/2n} \rightarrow a_i \ ,
\fqr
at the cost of doubling the unknowns from $\gamma_i$ to $a_i$.  There
is a trivial solution consisting of $a_i=1$ and $\gamma_i=0$.  The
equations are described by the numbers $(1-2n)^p$ in the general case,
and they may be analytically continued to $a_i$ the actual value.

The zeros of the Riemann function may be found by solving the infinite
number of equations in the infinite unknowns, for the true values of
$a_i$.  A methodology for obtaining the solutions to these polynomial
equations is described in \cite{ChalmersPoly}.

The polynomial equations with general $a_i$ are used to describe an
infinite dimensional, but highly symmetric (pseudo-pfaffian) algebraic
variety.  The metric on this space is used to find the allowed solutions
in $\gamma_j$ from the trivial solution $\gamma_j=0$ via geodesic flow;
this requires solving an infinite number of coupled second order
differential equations.  The presence of an infinite number of geodesic
components is simplified by the presence of few free parameters, i.e.
with the $a_i$ true there are no constants.

Label the space pertinent to \rf{polynomial} as $M_{P_c(z_i)}$ and its
standard Riemannian metric as $g_{\mu\nu}$.  Its K\"ahler so that both
$g_{\mu\nu}=\partial_{\mu}\partial{\nu} \ln\phi(z_i,{\bar z}_i)$ (in
terms of $z$ and ${\bar z}$, $g=g_{i{\bar j}}$) and its Christoffel
connection is derived as $\Gamma_{\rho,\mu\nu}= 1/2 \partial_\rho 
\partial_\mu\partial_\nu\ln\phi(z_i,{\bar z}_i)$ hold.

With the original point A to the solution of $P_c(z_i)=0$, i.e.
$\gamma_j=0$, a geodesic flow equation is given from point A to point B;
then $a_i$ are taken to the actual values.  The geodesic equation is,
with the coordinates $x=(z_i,{\bar z}_i)$,

\bqr
 {d^2 x^\rho \over d\tau^2} + \Gamma^{\rho,\mu\nu}
    {d x_\mu\over d\tau} {d x_\nu\over d\tau} = 0
 \label{geodesic}
\fqr
with,

\bqr
\Gamma_{\rho,\mu\nu}= 1/2 \partial_\rho \partial_\mu\partial_\nu
\ln\phi(x_\mu,{\bar x}_\nu) \ .
\label{Christoffel}
\fqr
The coordinates $x$ contain both the holomorphic and anti-holomorphic
components describing the geometry.  The flow from the original point 
A, such as at $a_i=0,\gamma_i=0$, to the points $B_j$ at which $a_i$ 
assumes its actual value, determine the constants $\gamma_i$.  These 
constants in turn determine the zeros $\zeta(\rho_i)=0$.  

\vskip .2in 

\noindent {\it Differential Approach}

The second approach involves solving a differential equation in $s$
that the zeta function satisfies.  The zeta function has a definition
in terms of

\bqr
\zeta(s)= {\pi^{s/2}/2} \Gamma(s/2)
  \int^\infty_0 \dt t^{s/2-1} \left[ \Theta(it)-1 \right] \ ,
\fqr
with the theta function defined as

\bqr
\Theta(it)= \sum_n e^{-n^2\pi it} \ .
\fqr
This form of the zeta function shows that the zeta function satisfies
the series of differential equations for integer $n$,

\bqr
{d^2 \zeta(\ln s)\over d s^2} = \zeta(\ln s + n) \ .
\label{zetadifferential}
\fqr
The form in \rf{zetadifferential} is not commonly reported in the
literature; the zeta function appears to be periodic form but takes
values on the plane.  The values within the strip $0 \leq {\rm Re}
\ln s < 1$ dictate the values everywhere else in the plane via the
derivatives.

Expanding the zeta function around the point $\ln s = n$ gives
the differential equation,

\bqr
{d^2\zeta(\ln s)\over ds^2} = \sum_{j=0}^\infty {(-n)^j\over j!}
  \zeta^{(j)}(t) \vert_{t=\ln s}
= e^{-n \partial_{\ln s}} \zeta(s) \ .
\fqr
This is a differential equation with an arbitrary number of derivatives
in the zeta argument.  Setting $\zeta(s)=0$ is a constraint imposed on
the differential equation.

The differential equation may be solved by analyzing geodesic flow on 
manifolds parameterized by the variables $\partial_{\ln s} \zeta(s)$.  
The method is presented in \cite{ChalmersGeoDiff}.  The metrics and 
the geodesic flows are required to be known in order to determine the 
solutions to the equation, and in particular to find the sub-manifold 
when $\zeta(s)=0$ (the derivatives are not necessarily vanishing on this 
slice).  

\vskip .2in 
\noindent{\it Zeta Zeros to Prime Numbers} 

As is well known, the zeta zeros allows a derivation of the prime numbers 
\cite{Zeta}.  The counting function on the prime determination is

\bqr
\pi(x) = \sum_m^n J(x^{1/m}) \mu(m)/m  \ ,
\fqr
with the $\mu(m)$ defined as: $0$ when $m^2/p^2 \in Z$, $1$ if $m=\prod
p_{\sigma(j)}$ and number of factors is even, and $-1$ if $m=\prod
p_{\sigma(j)}$ and the number of factors is odd.

Define the $J$ function as,

\bqr
J(x)={\rm Li}(x) - \sum_{\rho} {\rm Li} (x^\rho) - \ln 2 +
 \int_x^\infty {dt\over t(t^2-1)\ln t}  \ ,
\fqr
with $\rho$ a member of the set of solutions to $\zeta(\rho)=$, i.e.
$\rho=1/\gamma_j$.  The $\pi(x)$ function counts the prime numbers,
i.e. it bumps at the prime $N_i$.

The derivation of the zeta zeros in this work follows from either the
transcendental solution to the infinite set of equations, or from solving
an affiliated differential equation.  Methods for both are presented.
The metrics on the algebraic varieties are required for the explicit
derivation.

As a final comment, the elliptic L-series defined by

\bqr
L(C,s) = \prod \bigl( 1 - a_p p^{-s} + p^{1-2s} \bigr)^{-1}
\fqr
with $a_p = p - N_p$ numbering the integer solutions to the curve
with integral coefficients,

\bqr
y^2 = x^3 + ax + b \quad {\rm mod} \quad p  \ ,
\fqr
are found via the same method as used to find the Riemann function
zeros, i.e. in \cite{ChalmersPoly}.  The hyperelliptic L-series may
also be treated.

\vfill\break

\end{document}